# Quantum Computation of the Electronic Structure of Some Prototype Solids

Naman Khandelwal[1], Nidhi Verma[1], Pooja Jamdagni[2] and Ashok Kumar[1]*

[1]*Department of Physics, Central University of Punjab, Bathinda, 151401, India*

[2]*Department of Computational Sciences, Central University of Punjab, Bathinda, 151401, India*

(August 11, 2025)

*Corresponding Author:  ashokphy@cup.edu.in

# Abstract

Over the last decade, researchers have been working to improve a crucial aspect of quantum computing to predict Hamiltonian energy of solids. Quantum algorithms such as Variational Quantum Eigensolver (VQE) and Variational Quantum Deflation (VQD) have been used to study the molecular systems. However, there is growing interest in adapting and applying these methods to periodic solid-state materials. In this work, we have integrated first-principles density functional theory with VQE and VQD algorithms and utilizing the Wannier Tight-Binding Hamiltonian (WTBH) method to predict the electronic characteristics of solids. We demonstrate that VQE and VQD algorithms can be used to accurately predict electronic characteristics in a variety of multi-component prototype solid-state materials such as – Silicon (semiconductor), Gold (metallic), Boron Nitrile (insulator), Graphene (semi-metal). Efficient SU2 performs well among all the predefined ansatz used in the study. COBYLA is the fastest optimizer among the classical optimizers with minimum number of iterations for convergence. Results of noise models help to understand the band structure when calculated on real quantum hardware. As quantum hardware advances, our method stands as a prototype for future quantum simulations of materials pushing us closer to autonomous quantum discovery engines.

# 1. INTRODUCTION

Quantum computation of the electronic structure of solids is expected to be a future alternative to the computation based on classical computers[1,2]. Even with just a handful of qubits, quantum computers have the potential to surpass classical systems with larger number of atoms [3-5]. A crucial aspect of quantum computing in solid state systems is to predict energy levels of given Hamiltonian [6,7]. Researchers have been working for improving the estimation of Hamiltonian energy over the last decade by applying a variety of quantum techniques[8] such as variational quantum deflation (VQD) [9], variational quantum eigensolver (VQE) [10], and quantum phase estimation (QPE)[11]. These techniques help in the effectively calculating the energy states along with the improvement in the values of these states by augmenting additional methods such as quantum amplitude estimation (QAE) [12], quantum equation of motion(qEOM) [13], and quantum subspace expansion (QSE) [14]. Also, the quantum approximate optimization algorithm (QAQA) [15], quantum annealing [16], and the witness-assisted variational eigenspectra solver [17] methods have been used to reduce time of calculating energy states and accelerate the energy convergence. These techniques have been primarily used for the prediction of ground and excited states of Hamiltonian. However, their uses have been limited to molecular systems such as $BeH_2$, LiH, and $H_2$ [18-20].

Electronic structure calculations for solid-state systems are complex due to their periodic nature as quantum computation is mainly applied to the molecular systems. The electronic band structure of solids is important because it governs their electronic, optical, and thermal properties. Understanding band structure is thus critical in solid state physics research for designing and optimizing current technologies. Solid-state simulations are highly important to develop the design of superconductors, low-dimensional materials, and topological systems, which can ultimately improve noisy intermediate-scale quantum (NISQ) technologies [5,21,22].

VQE is acknowledged as an effective approach for calculating a Hamiltonian's ground state utilizing quantum mechanical principles in quantum computers. VQE involves developing an ansatz with configurable parameters and iteratively refining the parameters to reduce the Hamiltonian's expectation value with arbitrary accuracy [23]. Classical computers can develop Hamiltonian terms (decomposition in terms of Pauli matrices) as well as updating adjustable parameters during optimization. Quantum computers works on quantum states constructed using ansatz parameters and measuring interaction terms [24]. Researchers have been working to enhance the VQE technique for determining Hamiltonian energy levels beyond the ground state [9,25,26]. Due to the limitations of VQE algorithm in calculating higher energy states, new algorithms have been developed. VQD is a popular method for calculating higher energy levels which work on overlap estimation between the states and remove previously discovered eigenstates, allowing for the computation of excited-state energies and degeneracies. It has been noted that VQD uses the same number of qubits as that of VQE, and also VQE has restricted control of errors and enhancement in the accuracy [9].

VQD has been effectively used to determine higher energy levels in a variety of molecular systems [9,27] but still challenging for calculating periodic solid state systems. Integrating VQD technique with Wannier Tight-Binding Hamiltonian (WTBH) can be a promising approach for calculating energy bands of solids [28,29]. Wannier functions provide a comprehensive and orthonormal basis that bridges the gap between the delocalized plane-wave formalism commonly employed in electronic structure calculations, and a localized atomic orbital framework, which describes the chemical bonding more accurately [30,31]. WTBHs offer a computationally efficient way to investigate the electronic characteristics of solids. The Hamiltonian produced by WTBHs [32-34] are smaller, ranging from tens to hundreds of orbitals, making them ideal for noisy intermediate-scale quantum (NISQ) devices while plane-wave basis sets, which generate Hamiltonians that are too large to represent quantum hardware.

Recently, electronic structure of solids has been investigated using quantum algorithms[28,35]. Particularly, Cerasoli et. al. [35] have constructed Hamiltonians manually and the values of hopping parameters and energies were directly taken to construct the Hamiltonian matrices. However, limited data of hopping parameters and on-site energies make it difficult to create Hamiltonian matrix for variety of solid state systems [36]. The approach by Choudhary et. al, [28] relies on manually chosen materials based on JARVIS IDs and a simplified python script for the calculation of excited states of few solid state systems that can be applied on materials that have Hamiltonian in form of $2^n \times 2^n$ where n is the size of Hamiltonian matrix.

Our approach systematically extracts electronic Hamiltonians from Wannier functions generated via density functional theory (DFT). This enhances the scalability and generality of our method. Our approach also uses the widely used Qiskit platform [37,38] for calculating excited state which enhances the accuracy result when interfacing with actual quantum hardware. Four different prototype solids namely silicon (semiconductor), gold (metal), graphene (semi-metal) and boron nitride (insulator) have been considered in the current study.

## 2. THEORETICAL AND COMPUTATIONAL METHODOLOGY

The electronic Hamiltonian has been obtained utilizing Wannier functions which are generated from the DFT implemented in the Vienna Ab initio Simulation Package[39]. The GGA approximation parameterized by Perdew-Burke-Ernzerhof (PBE) has been used to obtain exchange-correlation functionals while the PAW technique was used to study the exchange of information between core and valence electrons [40]. The electronic Hamiltonian using hybrid HSE06 and GW method has also been obtained for semiconductor solid. A plane wave cutoff energy of 400 eV has been utilized having an 8×8×8 Monkhorst-Pack k-point grid throughout the simulations. For the structural relaxations, an energy convergence of $10^{-5}$ eV has been fixed. These models serve as input for quantum variational algorithms, allowing for accurate electronic structure simulations on Aer simulators (State vector simulators)[41].

The Brillouin zone is sampled using an uniform Monkhorst-Pack grid [42], where Bloch states are evaluated at each grid point. In this discretized k-space setting, differential operators (gradient and Laplacian) are approximated through finite-difference methods:

$$\nabla_k f(k) = \sum_b \omega_b b [f(k+b) - f(k)] \qquad (1)$$

$$\langle f(k) | \nabla_k^2 | f(k) \rangle = |\nabla_k f(k)|^2 = \sum_b w_b [f(k+b) - f(k)]^2 \qquad (2)$$

Here, each vector b corresponds to a displacement in k-space, and the magnitude b=|b| identifies a specific shell of neighbors. The weight factor $\omega_b$ is assigned to each shell and accounts for the symmetry and spacing of the grid, ensuring proper averaging and normalization in the finite-difference approximations of derivatives. Here k describes the wave vector and $\nabla_k, \nabla_k^2$ helps to convert spread functional to matrix elements. The set represents vectors that connect a given k-point on the Monkhorst-Pack mesh to its nearest neighboring points. These approximations are both accurate and efficient for functions that change smoothly with respect to the crystal momentum k.

The flow diagram of methodology used for band structure calculations is described in Figure 1. Maximally Localized wavefunctions (MLWFs) are defined using independent-particle within first principles calculations. Electronic characteristics of solids are described by single-particle Bloch wave functions($\psi_{nk}(r)$) in periodic potential ($u_{nk}(r)$) within the first Brillouin zone (BZ), where the wavefunctions are identified by their band number (n) and crystal momentum (k).

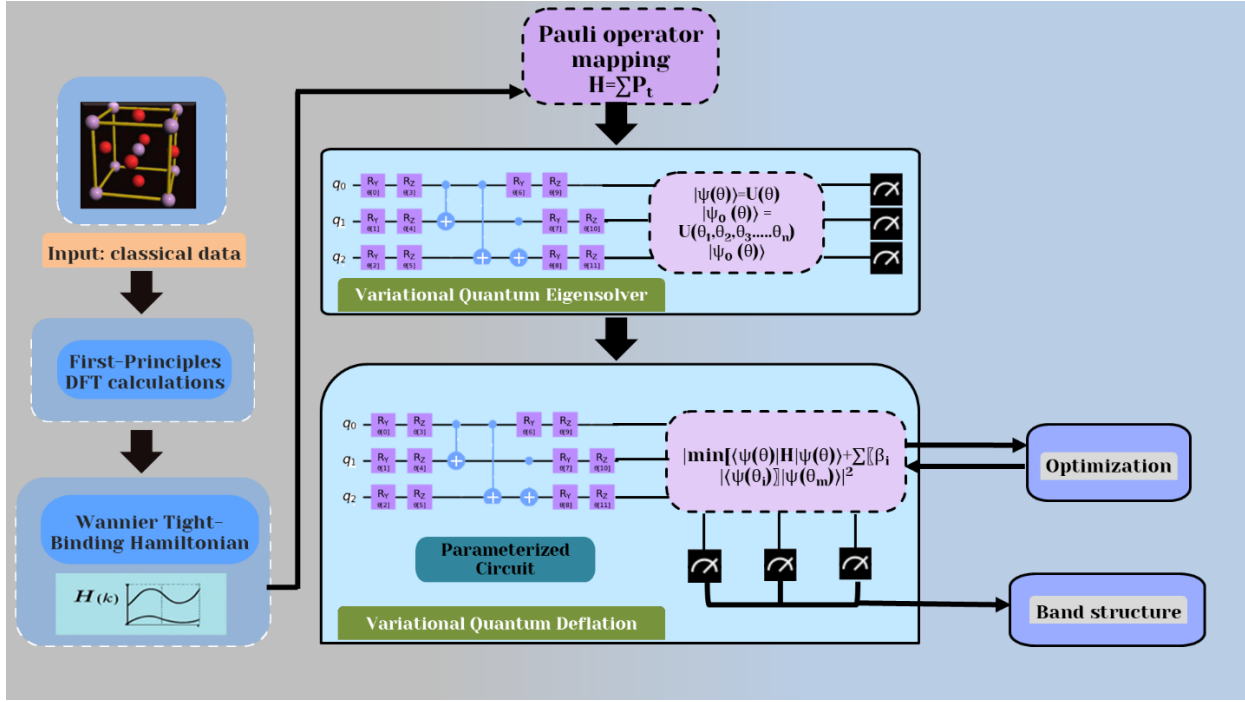

**Figure 1:** Flow diagram representing the quantum algorithmic steps for calculating electronic band structure of solids.

## 2.1 Hamiltonian generation using Wannier functions

The Wannier functions for periodic crystal can be derived from Bloch states by applying appropriate unitary transformations as [43] :

$$|w_{nR}\rangle = V \int_{BZ} \frac{dk}{(2\pi)^3} e^{-ik.R} \sum_{m=1}^{J} |w_{nR}\rangle U_{mnk} \qquad (3)$$

Here, V represents the volume of the unit cell, and $U_{mnk}$ are unitary matrices that combine Bloch states at each k-point. These matrices encapsulate the gauge freedom present in the Bloch functions, a freedom that is carried over to the corresponding Wannier functions. The Marzari–Vanderbilt (MV) strategy which involves selecting the set of unitary matrices U(k), minimizes the total quadratic spatial spread of the Wannier functions [44]. This spread is quantified by a spread function as:

$$\Omega = \Omega_I + \tilde{\Omega} \tag{4}$$

Where,

$$\Omega_I = \sum_n [\langle w_{n0}|r^2|w_{n0}\rangle - \sum_{mR}|\langle w_{mR}|r|w_{n0}\rangle|^2] \tag{5}$$

$$\tilde{\Omega} = \sum_n \sum_{mR \neq n0} |\langle w_{mR}|r|w_{n0}\rangle|^2 \tag{6}$$

The matrix elements of position operator in reciprocal space is represented as [45]:

$$\langle w_{nR}|r|w_{m0}\rangle = i\frac{V}{(2\pi)^3}\int e^{ik.R}\langle u_{nk}|\nabla_k|u_{mk}\rangle dk \tag{7}$$

$$\langle w_{nR}|r^2|w_{m0}\rangle = i\frac{V}{(2\pi)^3}\int e^{ik.R}\langle u_{nk}|\nabla_k^2|u_{mk}\rangle dk \tag{8}$$

These formulations allow the spread function to be expressed in terms of matrix elements involving the gradients of the Bloch functions with respect to crystal momentum $\nabla_k$ and their second derivatives $\nabla_k^2$. The two components of the spread function can be expressed as [46]:

$$\Omega_I = \frac{1}{N_{kp}}\sum_{k,b}\omega_b \sum_{m=1}^N [1 - \sum_{m \neq n}^N |M_{nn}^{(k,b)}|^2] \tag{9}$$

$$\tilde{\Omega} = \frac{1}{N_{kp}}\sum_{k,b}\omega_b [\sum_{n=1}^N (-Im \ln M_{nn}^{(k,b)} - b.\bar{r}_n)^2 + \sum_{m \neq n}^N |M_{nn}^{(k,b)}|^2] \tag{10}$$

Here $N_{kp}$ represents the number of k-points obtained using Monkhorst-Pack grid. $\bar{r}_n$ is the centre of the n$^{th}$ Wannier function represented as:

$$\bar{r}_n = \frac{1}{N_{kp}} \sum_{k,b} \omega_b \, b \, Im \, ln M_{nn}^{(k,b)} \qquad (11)$$

$\tilde{\Omega}$ is further decomposed into diagonal and off-diagonal parts as:

$$\tilde{\Omega} = \Omega_D + \Omega_{OD} \qquad (12)$$

Where,

$$\Omega_D = \sum_n \sum_{R \neq 0} |\langle w_{nR}|r|w_{n0}\rangle|^2 = \tilde{\Omega} = \frac{1}{N_{kp}} \sum_{k,b} \omega_b \left[\sum_{n=1}^{N}(-Im \, ln \, M_{nn}^{(k,b)} - b.\bar{r}_n)^2\right] \qquad (13)$$

$$\Omega_{OD} = \sum_{m \neq n} \sum_R |\langle w_{mR}|r|w_{n0}\rangle|^2 = \frac{1}{N_{kp}} \sum_{k,b} \omega_b \sum_{n=1}^{N} \sum_{m \neq n}^{N} |M_{nn}^{(k,b)}|^2 \qquad (14)$$

After calculating these elements, we can create Hamiltonian in MLWF's [46]. We find the Hamiltonian rotation of the Bloch states basis as:

$$H^{(W)}(k) = (U^{(k)})^\dagger (U^{dis(k)})^\dagger H(k) U^{dis(k)} U^{(k)} \qquad (15)$$

Where,

$$H_{nm}(k) = \varepsilon_{nk} \delta_{nm} \qquad (16)$$

$$H_{nm}(R) = \frac{1}{N_0} \sum_k e^{-ik.R} H_{nm}^{(W)}(k) \qquad (17)$$

The elements $H_{nm}(R)$ can be interpreted as the Hamiltonian matrix elements between MLWF's centered at lattice sites separated by the vector R. Owing to the strong spatial localization of the MLWF's, these matrix elements decay rapidly as the distance |R| increases. This localization property forms the foundation for applying a Slater-Koster-like interpolation scheme [47], which enables efficient reconstruction of the Hamiltonian on a much denser grid of k-points within the original Bloch representation. This is achieved by performing an inverse Fourier transform of the real-space Hamiltonian elements as:

$$H_{nm}(k') = \sum_R e^{ik'\cdot R} H_{nm}(R) \qquad (18)$$

After getting Hamiltonian H(k), it has to be mapped on the system of qubits.

**2.2 Pauli operators for qubit operations**

In practical quantum computing, qubit operations are performed using a set of Pauli matrices: X, Y, Z, and I, where I is the 2×2 identity matrix[38], where an Hermitian operator is represented as a linear combination of these Pauli matrices, forming a complete basis for matrices of size $N=2^n$, with $n=\log_2 N$. This basis is constructed by taking tensor products of the Pauli matrices across multiple qubits as:

$$\{\hat{\sigma}\}_n = [\text{I, X, Y, Z}]^{\otimes n} \qquad (19)$$

Decomposed Hamiltonian is than expressed as:

$$\hat{H}_k = \sum_{i=1}^{4^n} c_{ik} \hat{\sigma}_i \qquad (20)$$

Here, the set $\{\hat{\sigma}\}_n$ represents the complete set of $4^n$ Pauli basis matrices formed from tensor products over n qubits, and $(c_k)_n$ denotes the corresponding set of complex coefficients. These coefficients constitute the spectral decomposition of the Hermitian matrix with trace of the product defined as:

$$Tr(\hat{\sigma}_i^\dagger \hat{\sigma}_j) = 2^n \delta_{ij} \qquad (21)$$

This orthogonality allows each coefficient $c_k$ to be easily calculated using:

$$c_i = \frac{Tr(H_k^\dagger \hat{\sigma}_i)}{2^n} \qquad (22)$$

where H is the Hermitian matrix being decomposed. As a result, a linear combination of the $4^n$ Pauli basis matrices form a Hamiltonian with each term weighted by a corresponding coefficient.

## 2.3 Variational Quantum Eigensolver (VQE)

VQE is a prominent hybrid quantum-classical algorithm designed to find the minimum or maximum eigenvalue of an operator with relatively low quantum resource requirements[40]. It achieves this by alternating between quantum measurements and classical optimization routines [10,20,24]. The method is grounded in the Rayleigh-Ritz variational principle, which asserts that the ground state energy of a system can be approximated by minimizing the expectation value of the Hamiltonian over a trial wavefunction, parameterized by a set of variables. In a typical implementation, the preparation of quantum states and the measurement of expectation values are carried out on quantum hardware, while the parameter optimization is handled by classical algorithms.

The algorithm for estimating the ground state follows three key steps: Ansatz Preparation, Energy Measurement and Classical Optimization, as described in Figure S1, Supporting Information. In the first step, a parameterized quantum circuit $\hat{V}(\theta)$, known as the variational form or ansatz, is constructed. This circuit prepares a trial quantum state $|\psi(\theta)\rangle = \hat{V}(0)$, where $|0\rangle$ is the initialized reference state of all qubits. In the next step, expectation value of the Hamiltonian $\hat{H}_k$, is measured with respect to the parameterized state, yielding the energy $E(\theta) = \langle\psi(\theta)|\hat{H}_k|\psi(\theta)\rangle$. The Hamiltonian is expressed as a weighted sum of Pauli strings:

$$\hat{H}_k = \sum_{i=1}^{4^n} c_{ik} \hat{\sigma}_i \qquad (23)$$

where $\langle\sigma_i\rangle$ are tensor products of Pauli matrices and $c_{ik}$ are real coefficients. The wavefunction is measured in the Pauli basis to obtain $\langle\sigma_i\rangle$, which are then combined to compute the expectation value $\langle\hat{H}_k\rangle$. As quantum measurements are inherently probabilistic, multiple repetitions are needed to achieve desired accuracy. In the last step, a classical optimizer iteratively updates the parameter θ to minimize the energy function E(θ). The minimum value obtained corresponding to ground state energy is:

$$\varepsilon_0 = \langle \psi(\theta_{min}) | \hat{H}_k | \psi(\theta_{min}) \rangle \tag{24}$$

where $\theta_{min}$ are the optimal parameters yielding the lowest energy.

## 2.4 Variational Quantum Deflation (VQD)

The modified Hamiltonian of VQD algorithm for higher energy states is represented as [48]:

$$\hat{H}_1(m) = \hat{H}(m) + \beta_0 |\psi(\theta_0)\rangle\langle\psi(\theta_0)| \tag{25}$$

Where, the expectation value of Hamiltonian is given as:

$$\langle\psi(\theta)|\hat{H}_1(m)|\psi(\theta)\rangle = \langle\psi(\theta)|\hat{H}(m)|\psi(\theta)\rangle + \beta_0|\langle\psi(\theta)|\psi(\theta_0)\rangle|^2 \tag{26}$$

Subsequently, the first excited energy band is determined as:

$$E_1(m) = \langle\psi(\theta_1)|\hat{H}(m)|\psi(\theta_1)\rangle. \tag{27}$$

The parameter $\beta_0$ should be chosen large enough to act as a penalty term, enforcing orthogonality between the current trial state $|\psi(\theta_1)\rangle$ and the previously optimized state $|\psi(\theta_0)\rangle$. The $n^{th}$ excited energy band is determined by altering the original Hamiltonian as:

$$\hat{H}_n(m) = \hat{H}(m) + \sum_{j=0}^{n-1} \beta_j |\psi(\theta_j)\rangle\langle\psi(\theta_j)| \tag{28}$$

The expectation value is given by:

$$\langle\psi(\theta)|\hat{H}_n(m)|\psi(\theta)\rangle = \langle\psi(\theta)|\hat{H}(m)|\psi(\theta)\rangle + \sum_{j=0}^{n-1} \beta_j|\langle\psi(\theta)|\psi(\theta_j)\rangle|^2 \tag{29}$$

The $n^{th}$ excited energy band is obtained by computing $E_n(m) = \langle\psi(\theta_n|\hat{H}(m)|\psi(\theta_n)\rangle$ at each k-point. To maintain orthogonality with the previously determined states, the parameters $\beta_j$ are

selected such that $\beta_j > E_n - E_j$. The implementation of VQD algorithm is summarized in Figure S2, Supporting Information. For a chosen initial set of parameters $\theta_0$ and a specific k-point, the Quantum Processing Unit (QPU) generates the trial wavefunction $|\psi(\theta_0)\rangle$ and carries out Z-basis measurements as dictated by the qubit Hamiltonian. This Hamiltonian acts like a guideline, detailing how the trial wavefunction should be measured to evaluate its expectation value. The Classical Processing Unit (CPU) repeatedly suggests new parameter values $\theta_i$ based on a classical optimization algorithm and invokes the QPU each time to compute the expectation value. This iterative process continues until either the maximum number of iterations, denoted as $N_{max}$, is reached or the convergence criteria are met. The result of this loop is a parameter set $\theta_0$ that minimizes the expectation value. Notably, this procedure is carried out iteratively for each energy eigenvalue n = 0, 1, ..., 8. The same steps are repeated for each next k-point. This methodology offers a novel route for addressing band structure, balancing computational cost and accuracy.

## 3. RESULTS AND DISCUSSION

We investigated the electronic band structure for a variety of representative systems, including Gold, Silicon, Graphene and hexagonal Boron nitride. For each material, relevant atomic orbitals are determined to create Hermitian Hamiltonian matrices, which are subsequently encoded into quantum circuits suited for short-term quantum simulations. While these matrices can be diagonalized classically, they act as essential parameters for evaluating quantum technologies. The Hamiltonians are written as weighted sums of Pauli operators using unitary transformations, and the VQE is used to calculate the ground state at each k-point in the Brillouin zone. The final quantum circuits, which are built with qubits ordered sequentially from top (qubit 0) to bottom, provide the quantum system's complete instruction set. These circuits are then customized using classical optimizers in the Qiskit framework to reduce the Hamiltonian's estimated value.

To evaluate the accuracy and efficiency of this quantum workflow, we implemented multiple ansatzes—including hardware-efficient, real-amplitude, Excited State and Two local as shown in Figure 2. Among these, EfficientSU2 has been constantly performs in terms of convergence speed, final energy accuracy, and circuit resource efficiency. Its structured design, which includes alternating layers of Ry, Rz rotations and CNOT entanglers connected linearly, enables efficient entanglement across all qubits while reducing circuit depth and gate count. In contrast, the Real Amplitudes, Excited state and Two Local ansatzes either lack sufficient expressibility or incur additional overhead from deeper circuits and redundant parameters. Note that we adopted EfficientSU2 as the default circuit model in the calculations of electronic band structure. To ensure generalizability, we used a fixed depth with five repetition layers for each circuit across all prototype solids. For better circuit design, we also investigated many classical optimizers including COBYLA [49], L-BFGS-B [50], SLSQP, CG [51], and SPSA [52]—that are critical for optimizing the expectation value in the VQE loop. Each optimizer has distinct advantages: COBYLA shines in low-dimensional situations via its derivative-free linear approximations, whereas L-BFGS-B relies on quasi-Newton updates, making it appropriate for systems with large parameter spaces and bound constraints. The convergence behavior of these optimizers was observed during simulations of various solids at different k-points. COBYLA performs exceptionally well in low- to moderate-dimensional parameter spaces due to its use of derivative-free, trust-region-based linear approximations, which are robust in navigating rugged or noisy optimization landscapes common in variational quantum algorithms. As can be seen in Figure 3, COBYLA has the fastest convergence, whereas CG requires more iterations to achieve stability. SLSQP also performed well, converging at a similar rate to COBYLA. Once convergence was reached, the optimized parameters were utilized to calculate the system's ground-state energy as well as its quantum state. Further all calculations are carried out using COBYLA optimizer due to its best performance among these.

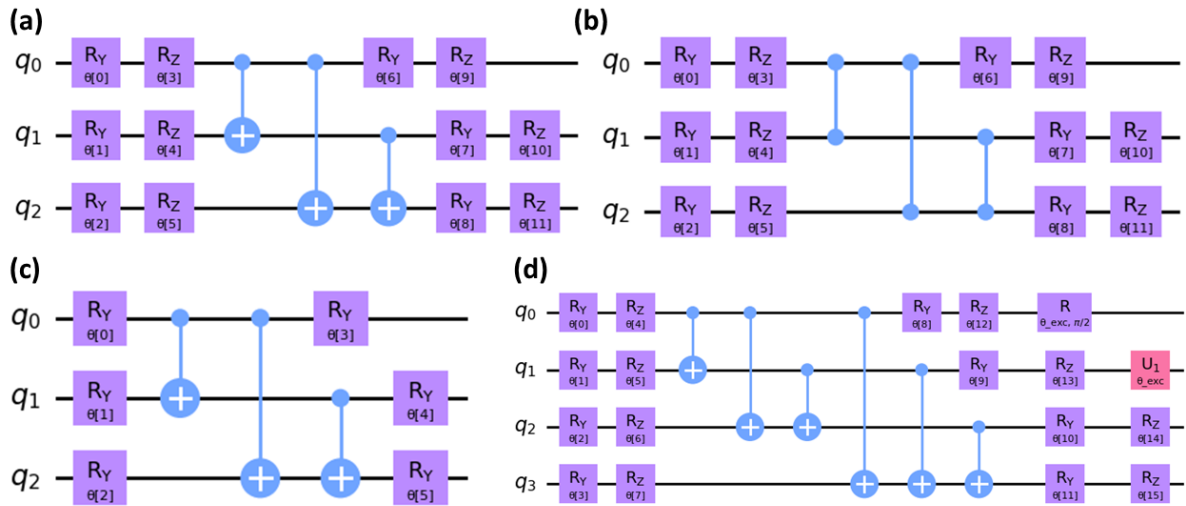

**Figure 2:** A collection of quantum circuit models with four key ansatzes: (a) EfficientSU2 (b) Two Local (c) Real Amplitudes (d) Excited state. These circuits incorporate parameterized rotation gates (RY and $R_Z$), denoted by angles θ.

After determining ground-state energy with the VQE, excited states were calculated using the VQD approach. This method was performed for all relevant k-points in the Brillouin zone to create the electronic band structures of the chosen materials. The simulations were carried out utilizing IBM's state vector simulator backend to confirm the approach, VQE-VQD findings were compared to classical reference eigenvalues produced using NumPy-based diagonalization [53], which repeatedly demonstrated high agreement between quantum and classical predictions. VQE is used for calculating lowest band energies which are carried out as a reference for calculating the higher band energies using the VQD algorithm.

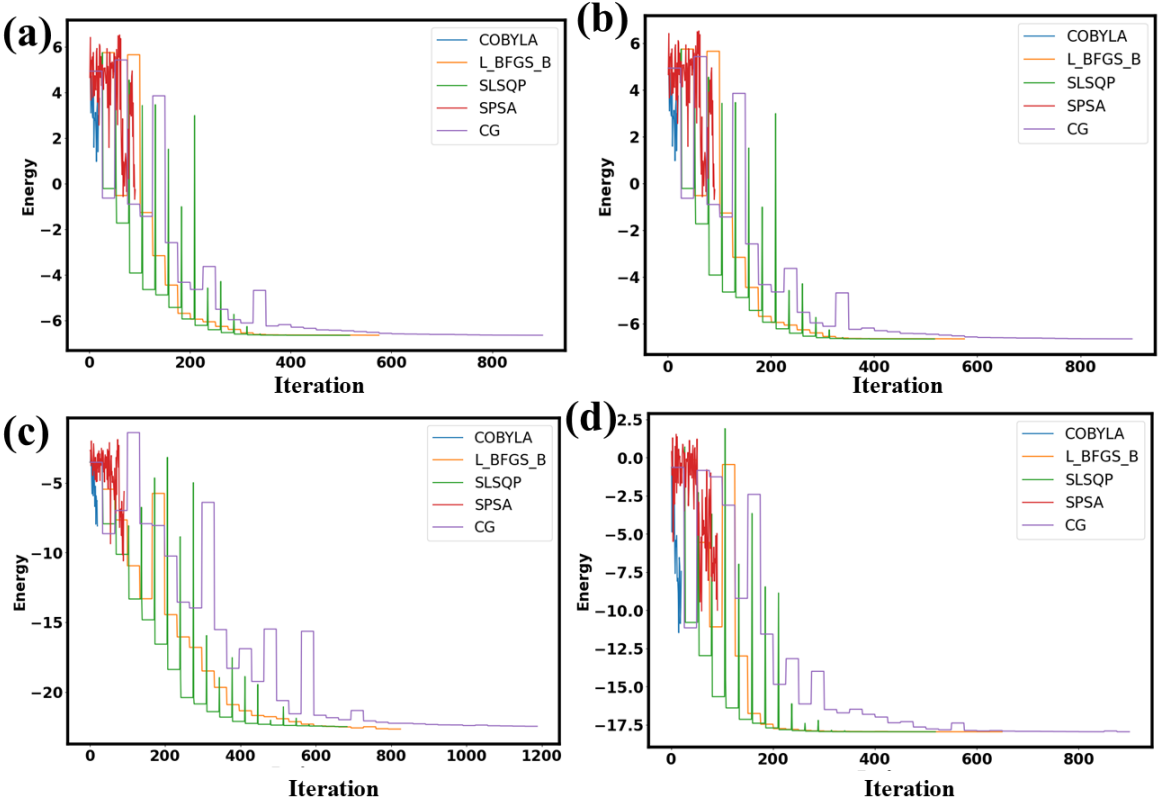

**Figure 3:** VQE convergence behavior for (a) Silicon (b) Gold (c) Graphene (d) Boron Nitride electronic WTBH at the Gamma point using different classical optimizers.

Quantum computed band structures for various prototype solids are compared to classical reference calculations derived from density functional theory (DFT) as shown in Figure 4. This demonstrates that WTBH, when paired with quantum algorithms like VQE and VQD, can accurately reproduce electronic band structures. One of the prototypes solid, Silicon, is also investigated beyond density functional level of theory utilizing hybrid HSE06 functional and GW method as these approaches accurately predicts the band gap of semiconductors. We have calculated the bandgap of 0.6 eV using PBE functional while the bandgap of 1.2eV and 1.6 eV have been evaluated using HSE06 hybrid and GW level of theory as depicted from the electronic band structure in Figure 5.

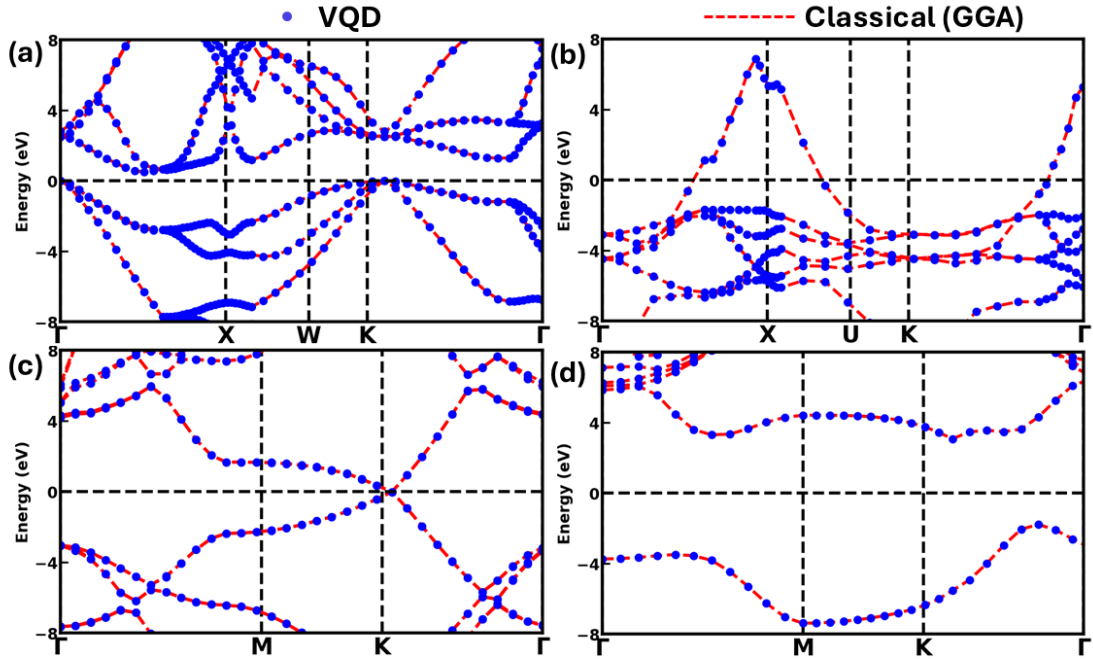

**Figure 4:** Electronic band structure calculated from classical (GGA) and VQD algorithms for (a) Silicon (b) Gold (c) Graphene (d) Boron nitride.

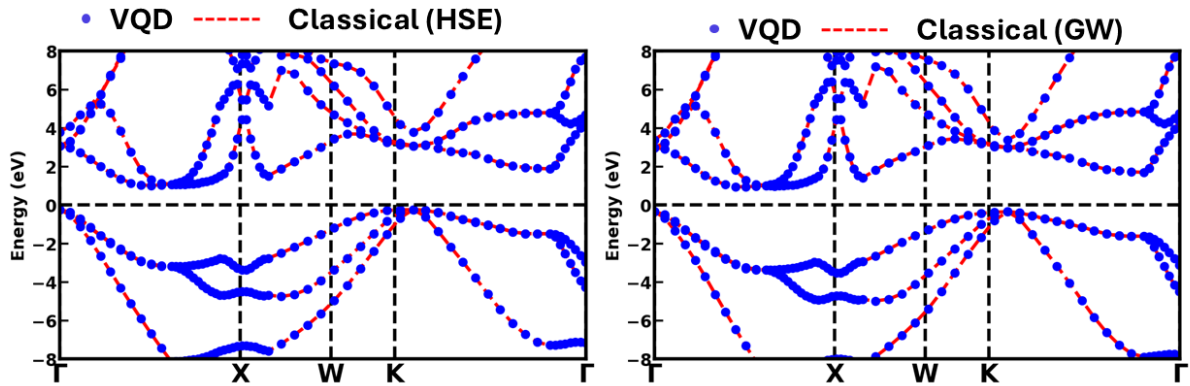

**Figure 5:** The comparison of the electronic band structure using VQD algorithm with hybrid HSE06 functional and GW-level of theory in classical computation for Silicon.

The selection of quantum circuit structures for variational quantum algorithms is usually problem-dependent, driven by both expressibility and hardware compatibility [54]. In this study, we used four different ansatzes and implemented to calculate band structure for all prototype solids as shown in Figure S3, Supporting Information. Among the four, EfficientSU2 consistently delivered more accurate energy predictions across all systems, albeit with increased gate complexity and parameter count.

Although a few ansatz layers (typically 2–3) are sufficient to capture the overall electronic structure of materials, accurately estimating the band gap often requires increasing the ansatz depth to achieve higher precision. Selecting the optimal ansatz and depth manually can be effective, developing an automated, system-aware ansatz selection framework would greatly enhance simulation efficiency. This comprehensive benchmarking framework not only validates the effectiveness of variational quantum algorithms for practical material simulations but also highlights how ansatz architecture and optimizer selection affect convergence, energy accuracy, and overall circuit performance across different electronic systems.

To test possible quantum circuits under practical situations, noise models were used to imitate real quantum hardware behavior. Noise models are errors implemented to every gate that are connected to ansatz circuit. The study investigated how the EfficientSU2, Two Local, and Real Amplitudes ansatzes responded to defects inherent in near-term quantum devices by accounting for noise effects such as gate failures, decoherence, and readout inaccuracies. These noisy simulations were run with Qiskit's noise module, which simulates backend-specific error characteristics found in actual IBM quantum computers. The presence of noise significantly influenced the convergence rate and precision of computed eigenvalues as shown in Figure S4, Supporting Information.

In actual quantum simulations, particularly on noisy intermediate-scale quantum (NISQ) devices, decoherence, gate problems, and readout noise may significantly decrease the accuracy of variational algorithms like VQE and VQD. To simulate these defects in classical simulations, we use a noise model, which quantitatively characterizes the physical error processes that occur during quantum computation. A noise model contains parameters like $T_1$ (relaxation time) and $T_2$ (dephasing time) to simulate qubit decoherence, as well as gate error probability (e.g., depolarizing or Pauli errors) and readout assignment errors. Noise models are implemented into Qiskit's Aer [38] backend by adding stochastic error channels after each gate operation or measurement, thereby simulating real-device behavior. The inclusion of a noise model is critical for assessing the resilience of ansatz circuits, optimizers, and algorithms under realistic conditions, allowing for benchmarking and error mitigation measures prior to execution on real hardware. In our study, we evaluate our quantum circuits against such noise models to analyze the fidelity loss, convergence behavior, and energy correctness of the computed band structures, guaranteeing that the suggested approaches are still viable for future deployment on real quantum processors. This process ensures that algorithmic choices remain effective even under noise and helps identify optimization and mitigation strategies to enhance result accuracy without full-scale error correction.

The electronic band structure with noise-module is shown in Figure 6. While the accuracy with noise-module is lower than in ideal (noise-free) simulations, the circuits, particularly EfficientSU2 still produced reasonably accurate results, exhibiting resilience in noisy situations. These findings imply that such quantum circuits could be installed on real quantum hardware with proper error mitigation measures.

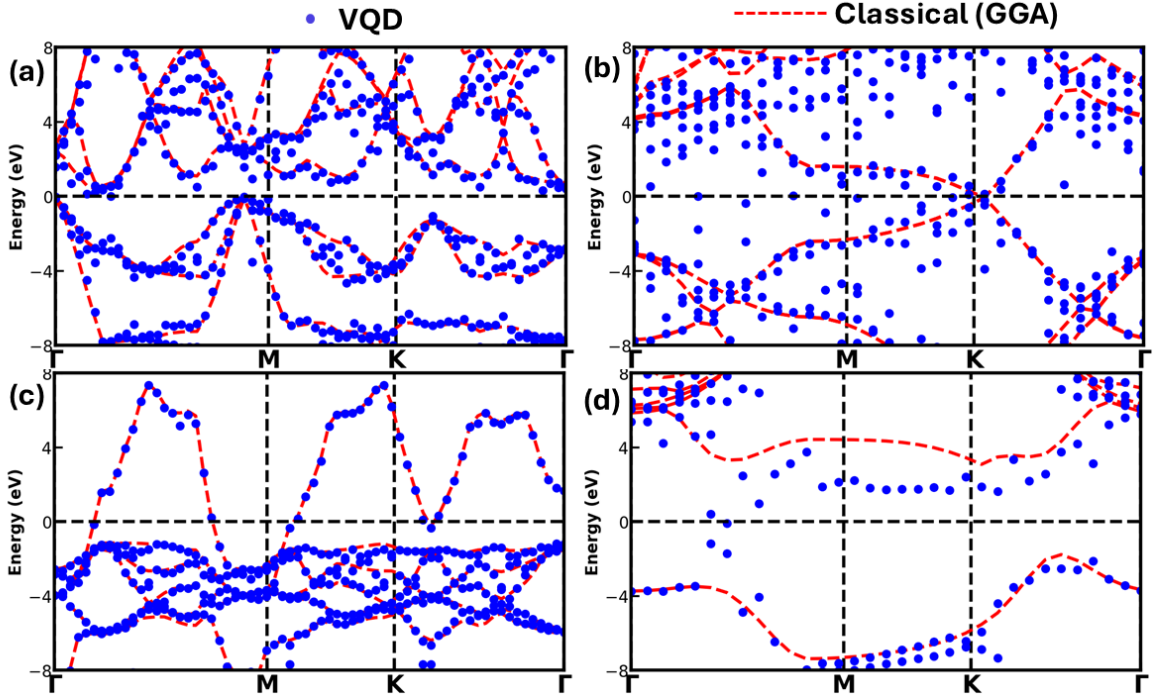

**Figure 6:** Computation of Band structures of (a) Silicon (b) Graphene (c) Gold (d) BN with Noise models.

Error mitigation strategies are critical in quantum computing because current NISQ systems produce noise due to decoherence, gate defects, and readout mistakes[55]. These errors build during circuit execution, resulting in misguided outputs, particularly in algorithms such as VQE or VQD, which depend on precise energy calculations. Because comprehensive quantum error correction is not yet possible, error mitigation offers realistic solutions to lessen the influence of noise without adding more qubits.

## 4. CONCLUSIONS

This study redefines the frontier of quantum simulation by extending excited-state quantum algorithms from molecules to realistic solid-state materials. Through the synergy of WTBH and VQD, we establish a pathway for momentum-resolved band structure computation on quantum devices. Our approach transforms dense *ab initio* data into qubit-ready Hamiltonians, enabling quantum circuits to explore the electronic landscape of materials with both efficiency and precision. This is more than a computational technique; it is a shift in how we conceptualize quantum materials research from classical approximations to quantum-native exploration. We performed all the calculations using some predefined ansatz from which Efficient SU2 performs better. We also performed calculations using various type of optimizers among them COBYLA performed with precise number of iterations for energy convergence. Different kinds of error mitigation techniques are required for capturing results more accurately. Our results not only validate the use of VQD for solid state Hamiltonians but also demonstrate the viability of this hybrid workflow for excited-state band structure calculations. This study represents one of the early efforts to bring together ab-initio solid-state physics tools and quantum computing in a single pipeline.

## ACKNOWLEDGEMENTS

Helpful discussions with Abdul Kalam, Kurudi V Vedavyasa, Ramandeep Singh and Sangeeta Meena are highly acknowledged. NV also acknowledges CSIR for the financial assistance in the form of Senior Research Fellowship.

**Data availability statement**

All data generated or analysed during this study are included in this article [and its supplementary information files].

**Declarations**

The authors declare no competing interests.